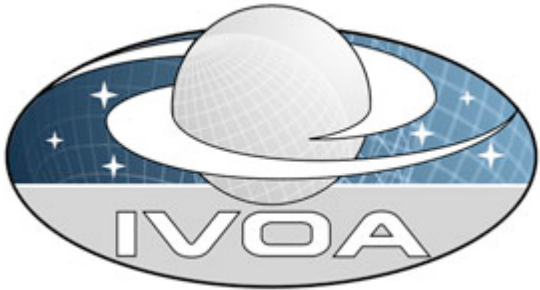

**International**

**Virtual**

**Observatory**

**Alliance**

# Simple Line Access Protocol

# Version 1.0

## *IVOA Recommendation 09 December 2010*

**Editor(s):**
**Pedro Osuna**
**Jesus Salgado**

**Author(s):**
Jesus Salgado
Pedro Osuna
Matteo Guainazzi
Isa Barbarisi
Marie-Lise Dubernet
Doug Tody

# Abstract

The **Simple Line Access Protocol (SLAP)** is an IVOA *Data Access protocol* which defines a protocol for retrieving spectral lines coming from various *Spectral Line Data Collections* through a uniform interface within the VO framework. These lines can be either observed or theoretical and will be typically used to identify emission or absorption features in astronomical spectra.

It makes use of the **Simple Spectral Line Data Model (SSLDM [1])** to characterize spectral lines through the use of **uTypes [14]**. Physical quantities of units are described by using the standard **Units DM [15]**.

SLAP services can be registered in an IVOA *Registry of Resources* using the **VOResource [12] Extension** standard, having a unique **ResourceIdentifier [13]** in the Registry.

The SLAP interface is meant to be reasonably simple to implement by service providers. A basic query will be done in a wavelength range for the different services. The service returns a list of spectral lines formatted as a VOTable. Thus, an implementation of the service may support additional search parameters (some which may be custom to that particular service) to more finely control the selection of spectral lines.

The specification also describes how the search on extra parameters has to be done, making use of the support provided by the **Simple Spectral Line Data Model (SSLDM [1])**

# Link to IVOA Architecture

The figure below shows where SLAP fits within the IVOA architecture:

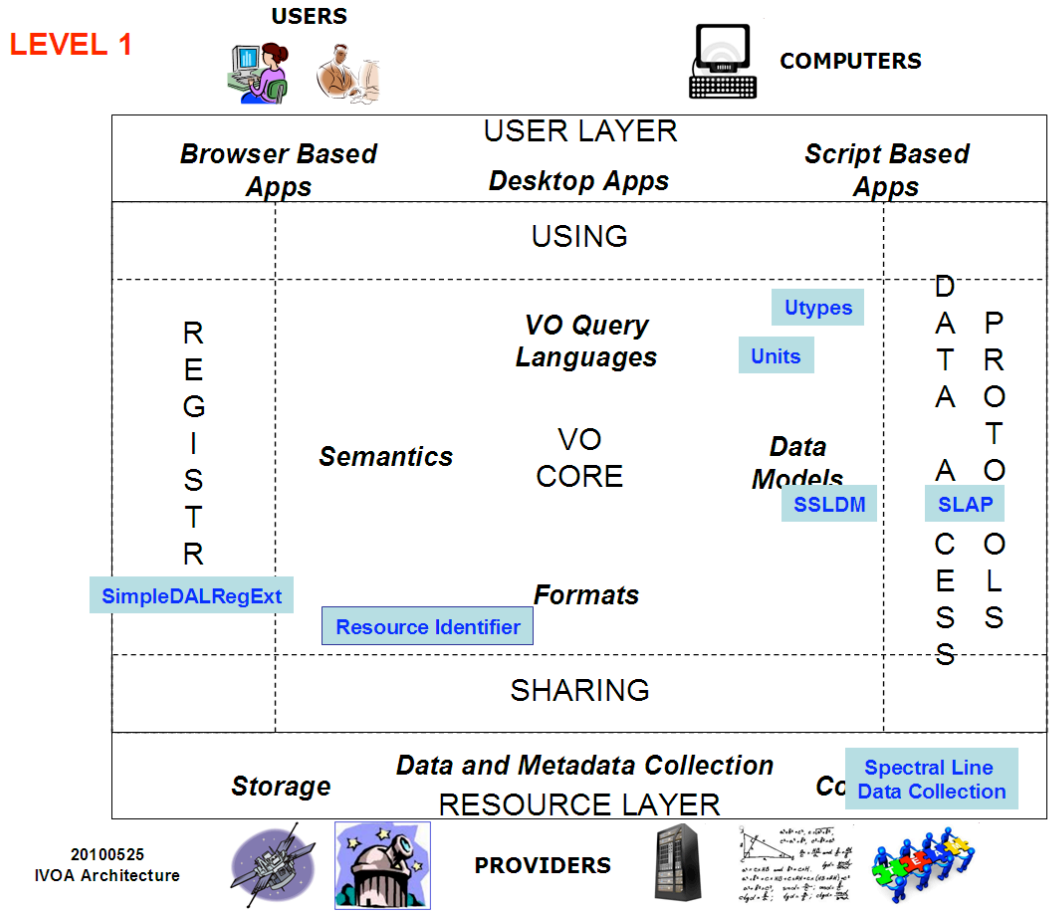

# Status of This Document

*This document has been produced by the IVOA **DAL** Working Group.*

*It has been reviewed by IVOA Members and other interested parties, and has been endorsed by the IVOA Executive Committee as an IVOA Recommendation. It is a stable document and may be used as reference material or cited as a normative reference from another document. IVOA's role in making the Recommendation is to draw attention to the specification and to promote its widespread deployment. This enhances the functionality and interoperability inside the Astronomical Community.*

*A list of current IVOA Recommendations and other technical documents can be found at http://www.ivoa.net/Documents*

# Acknowledgements

The authors acknowledge the comments from the DAL forum members and from the IVOA members in general.

# Change Log

| Change | Section | Date |
|---|---|---|
| Abstract updated removing tentative language | Abstract | 20091016 |
| Citation added to SSLDM | Abstract | 20091016 |
| Added paragraph clarifying must/should/may use | 2.1 | 20091016 |
| Last paragraph, section 3, wording improved | 3 | 20091016 |
| Input format, Base URL and Set of Constraints wording improved | 4.1, 4.1.1, 4.1.2 | 20091016 |
| Backwards compatibility comments deleted | 4.2.1 | 20091016 |
| Compulsory parameters section name changed to Required parameters | 4.2..2 | 20091016 |
| Use of score instead of sort | 4.2.3.3 | 20091016 |
| Custom query parameters instead of Free query parameters | 4.2.3 | 20091016 |
| Section 7 about registration updated removing registry specific details | 7 | 20091016 |
| Sections 5.1 and 5.1 condensed to 5.1 | 5 | 20091016 |
| Added extra citations | References | 20091016 |
| Abstract wording improved | Abstract | 20091016 |
| Consistency explanation with SSAP and SIAPv2 protocols added | 1 | 20091016 |
| Remark of parameters in query need | 4.1 | 20091016 |
| Although appendix on ranges is preserved, citation to SSAP done | 4.1.2 | 20091016 |
| Standard parameters section created to describe VERSION & REQUEST | 4.2 | 20091016 |
| Description of parameters section added | 4.2 | 20091016 |
| References to name spaces syntax updated | 5 & 6 | 20091016 |
| LaTeX use clarified | 6.1 | 20091016 |
| Quotes removed in process type example | 6.1 | 20091016 |

| Change Proposed by Recommended and Compulsory by Required | Appendix B and D | 20091016 |
|---|---|---|
| Typo ENEGY instead ENERGY corrected | 4.1 | 20090624 |
| Link to IVOA architecture | Abstract | 20090805 |

# Contents

# 1  Overview

This Simple Line Access Protocol (SLAP henceforth) specification makes use of the work done in the Simple Spectral Lines Data Model (SSLDM henceforth) definition, as the source of the abstract representation of a spectral line. All the information collected in this document will be used to homogenize the access to the different existing spectral line databases.

The approach followed by this protocol is the same one as the successful SIAP protocol. This document is intended to be a translation of the SIAP protocol specification for spectral lines access, diverging from it when the physics of the problem force us to do so. At the same time, some ideas of the access to theoretical spectra have been developed and used for Theoretical Spectral Line databases.

The SLAP interface has intentionally been made similar to that of SSAP and SIAP (v2.0) through the adoption of current IVOA Data Access Layer conventions, including:

- queries encoded as URLs,
- the use of standard input parameters, REQUEST and VERSION,
- the syntax for specifying numeric ranges and value lists,
- the use of VOTable for encoding search results,
- the mechanism for handling errors, and
- the retrieval of service metadata.

However, SIAP and SSAP protocols are two-step processes. In the first step, the VO client application requests metadata from the server. These metadata include links to images or spectra. In the second step, the VO client application requests the data from the server. In the particular case of Simple Line Access Services, these kind of services diverge from the SIAP and SSAP ones, as the SLAP service will be, in essence, a one-step process, i.e., only the first request for metadata is needed.

Even when no link to astronomical products is expected because of the nature of the service, the SLAP metadata could contain reference links to html pages, spectra files, spectral line profiles, etc.

## 2   Requirements for Compliance

The spectral line query web method **MUST** be supported as in section 4 below. Through this web method, clients search for spectral lines that match certain client-specified criteria. The response is a VOTable that describes the output spectral lines.

### 2.1   Compliance

The keywords **MUST**, **REQUIRED**, **SHOULD**, and **MAY** as used in this document are to be interpreted as described in RFC 2119 [34].

An implementation is compliant if it satisfies all the **MUST** or **REQUIRED** level requirements for the protocols it implements. An implementation that satisfies all the **MUST** or **REQUIRED** level and all the **SHOULD** level requirements for its protocols is said to be "unconditionally compliant"; one that satisfies all the **MUST** level requirements but not all the **SHOULD** level requirements for its protocols is said to be "conditionally compliant".

# 3 Spectral Line Service Types

It is assumed that compliant spectral line services fall into one of two categories.

1. Observational line databases. Lines observed and identified in real spectra collected by different instrument/projects.

2. Theoretical line databases. Servers containing theoretical spectral lines will be included in this group.

In both cases, the line description and the identification might be already present in a scientific publication, which could be used as a curation mechanism.

This document describes standard query parameters for SLAP services. Some SLAP services might make use of extra parameters, not cited in this document, to support additional filtering and selection.

However, the theoretical line database services could make use of extra parameters not cited in this document to filter out lines not expected to be identified in an observed spectrum or to score the output lines due to the application of physical models.

Examples:

For the Atomic spectral line database from CD-ROM 23 of R. L. Kurucz
> The absorption oscillator strength, log(gf), can be selected

For the NIST Atomic Spectra Database Lines
> As it is extracted from the Saha-LTE model, "*The level populations are calculated according to the Boltzmann distribution within each ion and Saha distribution between the ion stages. Thus, to calculate the spectrum from a single ion, e.g., C I, only $T_e$ is required, while for the spectrum from several ions of the same element (e.g., C I-V), $N_e$ must be defined as well.*"

At the same time, for observational spectral line databases, some project specific search parameters may be used.

Example:

ISO Astronomical Spectroscopy Database (IASD)
> The observation number parameter can be used to select only the lines observed during this ISO satellite observation.

Since it is awkward to compile all such extra parameters in a document such as this one, a general mechanism is described. As will be explained later, the discovery of these extra parameters by VO client applications or by the registry relies on an implementation of the format=metadata input parameter. The exact expected response when this parameter appears in the query will be discussed in point 5.

# 4  Spectral Line Query

The purpose of the spectral line query is to allow users/clients to search in a wavelength range for spectral lines. The most basic query parameters will be the minimum and maximum value for the wavelength range. Additional parameters may be used to refine the search or to model physical scenarios.

## 4.1  Input format

A URL with two parts is used to transmit the spectral line query as an HTTP GET request. These two parts are:

1. A base URL
2. A set of constraints

A legal SLAP URL should contain a base URL and at least the compulsory set of constraints (see next section).

### 4.1.1  Base URL

The service base URL has the form:

```
http://<server-address>/<path>? [<extra GET arg>&[...]]
```

Example:

```
http://esavo02.esac.esa.int:8080/slap/slap.jsp?
```

Note that when it contains extra GET arguments, the base URL ends in an ampersand, &; if there are no extra arguments, then it ends in a question mark, ?.

Every query to a given spectral line query service uses the same base URL.

### 4.1.2  Set of Constraints

Constraints expressed as a list of ampersand-delimited GET arguments, each of the form:

```
<name1>=<value1>[&<name2>=<value2>&…]
```

Example:

```
WAVELENGTH=5.1E-6/5.6E-6
```

The constraints represent the query parameters that can vary for each successive query.

A parameter value represents a range constraint when it matches the syntax for range values as defined in the SSAP protocol [8] section 8.7.2.

This range syntax will be used for the wavelength range and for other input parameters.

For example:

```
INITIAL_LEVEL_ENERGY =8.01E-18/1.6E-17
```

is equivalent to

```
INITIAL_LEVEL_ENERGY>=8.01E-18 AND INITIAL_LEVEL_ENERGY<=1.6E-17
```

See Appendix A for a summary of this syntax.

## *4.2  Input Parameters*

Parameters may appear in any order.  If the same parameter appears multiple times in a request, the operation is undefined (if alternate values for a parameter are desired, the range-list syntax may be used instead).  Parameter names are case-insensitive.  Parameter values are case-sensitive unless defined otherwise in the description of an individual parameter.

The following subsections define reserved parameters. In addition to the description of the functional role the parameter plays in constraining a query, each definition also includes, when applicable, a UCD [10] and/or UType that indicates the semantic meaning of the parameter. The UType names refer to those defined in the Simple Spectral Line Data Model [1].

### 4.2.1  Standard Parameters

All operations define the following standard parameters:

- **REQUEST**  A service **MUST** implement a REQUEST parameter that indicates which server operation is being invoked. The value shall be the name of one of the operations offered by the server.  It is an error to reference an unknown service operation. Valid values for parameter REQUEST in the case of a SLAP service are:
queryData, getCapabilities(reserved), getAvailabitity(reserved)

The response to a REQUEST=queryData operation would be a normal query response (see next sections). If the REQUEST parameter does not appear in the query, the default value is queryData (see example of use in next section).

The response to a REQUEST=getCapabilities operation would be a description of the service capabilities as described later in the Metadata Query section. For the time being, getCapabilities and getAvailability are only reserved but not fully defined in the present protocol version.

- **VERSION** A service **SHOULD** implement a VERSION parameter that indicates to the service the expected SLAP version to be used in the client-server communication.

  Valid values in the form:

| <major_version>.<minor_version>[.<patch_version>] |
| --- |

| **Examples:** |
| --- |
| 1.0 |
| 1.2.1 |

  If a SLAP client does not specify the version number in a request, the server assumes the highest standard version supported by the service, and no explicit version checking takes place. If the client specifies an explicit version number, and this does not match a version available from the service at level two, the service returns a version number mismatch error. The client can determine what versions of the protocol the service supports by a prior call to getCapabilities or via a registry query. Please refer to SSAP for version negotiation.

The values of both the REQUEST and VERSION parameters are case-insensitive.

A given service instance may support multiple versions of the SLAP interface, which includes both the input parameters and the query response with all of its complex metadata, and by default the service assumes the highest standard version which is implemented (access to any experimental versions supported by a service requires explicit specification of the version by the client). Explicit specification of the interface version assumed by the client is necessary to ensure against a runtime version mismatch, e.g., if the client caches the service endpoint but a newer version of the service is subsequently deployed. The client can omit the VERSION parameter to disable runtime version checking and default to the highest version standard interface implemented by the service.
All other request parameters are defined separately for each operation.

## 4.2.2 Required parameters

A service must support the input parameters described in this section. That means that the service must accept them as valid ones without raising an error, and the parameters must be properly used to constrain the query.

### 4.2.2.1 WAVELENGTH

The service **MUST** support the **WAVELENGTH** parameter, to specify the wavelength spectral range, to be specified in meters. This wavelength range will be interpreted as the wavelength in the vacuum of the transition originating in the line (ucd="em.wl" ;utype="**ssldm:Line.wavelength.value**").

For range definition, please refer to Appendix A.

As the units in the spectral line database could be different than meters, the service will need to translate from the selected units (meters) to the internal ones. The selection of one type of units (in this case the SI unit meters) will help to unify access to different spectral line databases, even when in some cases, the unit selected (meter) may not be the best one for the range on interest.

Example

To query for spectral lines in the wavelength range between 5.1 and 5.6 micrometers:

http://esavo02.esac.esa.es:8080/slap/jsp/slap.jsp?REQUEST=queryData&WAVELENGTH=5.1E-6/5.6E-6

## 4.2.3 Non-compulsory Parameters

The next list of non-compulsory parameters may be implemented on the server side:

### 4.2.3.1 CHEMICAL_ELEMENT

A service **MAY** have a search parameter called **CHEMICAL_ELEMENT**. This parameter would constrain the search to the chemical element selected. A list of different chemical elements could be queried using the comma separator as described in Appendix A:

(ucd="phys.atmol.element";utype="**ssldm:Line.initialElement.name"**)

### 4.2.3.2 Energy Level range

A service **MAY** accept constraints in the energy for the starting and final levels of the transition. For these parameters, only a range constraint complaint with Appendix A is allowed.

#### 4.2.3.2.1 INITIAL_LEVEL_ENERGY

A service **MAY** implement the parameter **INITIAL_LEVEL_ENERGY** to specify the minimum and maximum energy for the INITIAL level of the transition, to be expressed in Joules
(ucd=" phys.energy; phys.atmol.initial;phys.atmol.level";
          utype="**ssldm:Line.initialLevel.energy.value"**)

#### 4.2.3.2.2 FINAL_LEVEL_ENERGY

A service **MAY** implement a parameter to specify the minimum and maximum energy for the FINAL level of the transition, to be expressed in Joules
(ucd=" phys.energy; phys.atmol.final;phys.atmol.level";
          utype="**ssldm:Line.finalLevel.energy.value"**)

### 4.2.3.3 TEMPERATURE

A service **MAY** implement the parameter **TEMPERATURE** to specify the expected temperature of the object, to be specified in Kelvin. This parameter would be used (in particular for theoretical spectral line databases) to sort the lines in the output using physical models.
(ucd="phys.temperature";utype="**ssldm:Line.environment.temperature.value**")

### 4.2.3.4 EINSTEIN_A

A service **MAY** implement a parameter **EINSTEIN_A** to accept constraints in the transition probability by specifying the minimum and maximum Einstein A, defined as the probability per unit time $s^{-1}$ for spontaneous emission in a bound-bound transition, to be specified in $s^{-1}$.
(ucd=" phys.atmol.transProb"; utype="**ssldm:Line.einsteinA.value**").
For range definition, please refer to Appendix A.

### 4.2.3.5 Physical Process Parameters

It is possible to discriminate the result lines per physical process that originate or modify the line. To do that, and in line with SSLDM, the following input parameters could be added.

#### 4.2.3.5.1 PROCESS_TYPE

A service **MAY** implement the parameter **PROCESS_TYPE** to specify the physical process type responsible for the generation of the line or for the modification of its physical properties (utype="**ssldm:Process.type**"). Valid values are: "Matter-radiation interaction", 'Matter-matter interaction", "Energy shift", "Broadening".

#### *4.2.3.5.2 PROCESS_NAME*

A service **MAY** implement the parameter **PROCESS_NAME** to specify the physical process responsible for the generation of the line or for the modification of its physical properties (utype="**ssldm:Process.name**").   There is a great variety of possible values, so this input parameter would need to make use of the FORMAT=METADATA discovery query, as described in next section.

Some possible values are: "Photoionization", "Collisional excitation", "Gravitational redshift", "Stark broadening", "Resonance broadening", "Van der Waals broadening", etc

### 4.2.4  Custom query parameters

As we saw in Section 3, there is a need to have a general mechanism for free query parameters to filter out or sort the table result.

Both for the non-compulsory parameters and/or for the free ones, client applications can discover whether a particular parameter is implemented through the FORMAT=METADATA response until the getCapabilities operation is fully defined. The exact format of this response will be defined in the next section.

Using this information, a VO client would be able to dynamically construct a form, where this information could be entered.

## 5  Service Metadata

Simple Spectral Line services advertise their availability by registering with a registry service, which may provide the mechanisms to fully characterize the service, including its non-compulsory and additional parameters. Alternately, the FORMAT=METADATA parameter or REQUEST=getCapabilities can be used to ascertain the service's full capabilities.

A compliant simple spectral line service **MUST** support spectral line queries with FORMAT=METADATA or REQUEST=getCapabilities, used to query the service metadata; only metadata is returned by the service in this case.  The response to this query advertises the input parameters the service supports.

Please note that until a proper definition of the getCapabilities is provided to supersede current FORMAT=METADATA operations, the service could use the described FORMAT=METADATA response as a valid getCapabilities response.

The structure of the FORMAT=METADATA result will be a VOTable with the parameters supported by the service, using the VOTable <PARAM> tag. Every input parameter supported by the service should be listed as a PARAM element of the RESOURCE that normally contains the spectral line list table. Each param should have a name attribute in the form "INPUT:param_name", where param_name is the parameter name as it should appear in the query URL. For example, name="INPUT:WAVELENGTH" refers to the input parameter "WAVELENGTH".

All input parameters meant to be available to clients of the service must be listed as PARAM elements, including required parameter WAVELENGTH, non-compulsory parameters and free parameters specific to the service. The PARAM may contain a value attribute that should contain the default value that will be assumed if the parameter is not set in the query input URL. Implementers are encouraged to include, as children of the PARAMs, DESCRIPTION elements to describe the parameter and (where appropriate) VALUES elements to given allowed ranges or values.

At the same time, implementers are encouraged to include UCD elements in the PARAM tags to describe the physical meaning of one parameter, not always obvious in particular for physical models.

---

**Example:** The input parameter listing below from the Observed Spectral Lines database shows that in addition to supporting the required parameter (WAVELENGTH), it also supports the free parameter OBSNO.

```
<?xml version="1.0" encoding="UTF-8"?>
<VOTABLE xmlns:xsi="http://www.w3.org/2001/XMLSchema-instance"
xsi:noNamespaceSchemaLocation="xmlns:http://www.ivoa.net/xml/VOTable/VOTable-1.1.xsd"
xmlns:ssldm="http://www.ivoa.net/xml/SimpleSpectrumLineDM/SimpleSpectrumLineDM-
v1.0.xsd" version="1.0">

<RESOURCE type="results">
    <DESCRIPTION>IASD Simple Line Access Service</DESCRIPTION>
    <INFO name="QUERY_STATUS" value="OK"/>
    <PARAM name="INPUT:WAVELENGTH"
                            ucd="em.wl" utype=" ssldm:Line.wavelength.value"  value="30">
      <DESCRIPTION> Specify the wavelength spectral range. To be specified in meters. This
wavelength will be interpreted as the wavelength in the vacuum of the transition originating the
line
.      </DESCRIPTION>
    </PARAM>

    <PARAM name="INPUT:OBSNO" ucd="meta.id;obs">
      <DESCRIPTION> Specify the ISO observation number where this line has been identified
.      </DESCRIPTION>
    </PARAM>

    …………………………………..
```

# 6  Successful Output

The output returned by a SLAP service is a VOTable [6], an XML table format, returned with a MIME-type of "text/xml;content=x-votable". The table lists all the Spectral lines found in the server database that match the query constraints. The following requirements are placed on the contents of the table when the query successfully returns a list of spectral lines:

1. The VOTable **MUST** contain a RESOURCE element identified with the tag type="results" containing a single TABLE element which contains the results of the query. The VOTable is permitted to contain additional RESOURCE elements, but the usage of any such elements is not defined here. If multiple resources are present it is recommended that the query results be returned in the first resource element.
2. The RESOURCE element **MUST** contain an INFO with name="QUERY_STATUS". Its value attribute should be set to "OK" if the query executed successfully, regardless of whether any matching spectral lines were found.

**Examples:**
    <INFO name="QUERY_STATUS" value "OK">
    <INFO name="QUERY_STATUS" value="OK"> Successful Search </INFO>

3. Each table row represents a different spectral line available to the client.
4. Each row of the output VOTable **MUST** contain FIELDs.
5. Every FIELD **SHOULD** contain a utype reference to the Simple Spectral Line Data Model whenever possible.
6. A standard column **MUST** have a defined utype and a defined UCD as described in next section
7. A standard column could appear multiple times with different units. The way to uniquely identify one standard column is the following:
   - When a standard column can appear multiple times with the same utype but different units, the column is uniquely identified by its utype and unit.
   - Otherwise, a standard column is uniquely defined by its utype.
8. The VOTable **MUST** contain a reference to the SSLDM namespace

    xmlns:ssldm="http://www.ivoa.net/xml/SimpleSpectrumLineDM/SimpleSpectrumLineDM-v1.0.xsd"

9. The VOTable **MAY** contain references to other name spaces, like SLAP, Characterization, etc

## *6.1  Standard output fields*

- One field **MUST** have utype="**ssldm:Line.wavelength.value**", with datatype="double" and ucd="em.wl", containing the wavelength in vacuum of the transition originating the line in meters.
  It is allowed to have more than one wavelength field in different units in order to preserve the precision of the original value prior to unit conversion in the client. If this is the case, and to get backwards compatibility with already existing services, there **MUST** be one field with utype="**ssldm:Line.wavelength.value**" and unit="m" in the VOTable response. Other fields with the same utype should have a different value in the unit field descriptor.

- Exactly one field **MUST** have utype="**ssldm:Line.title**", with datatype="char" arraysize="*" and ucd="em.line", containing a small description identifying the line.

Note that this line title is only a short string representation to be used in the clients for display. There is no required syntax, but it is recommended that common species and transition notation be used when applicable.

| **Examples:** |
| --- |
| H I |
| N III 992.873 A |

In case of corrected but unidentified lines, some examples could be:

| **Examples:** |
| --- |
| M31 1001.784 A |
| 011910191 800.2 A |

- Exactly one field **SHOULD** have
  utype="**ssldm:Line.identificationStatus**" with datatype="char", arraysize="*" , containing the identification status of the line. Possible values are: unidentified, identified and uncorrected. Note that this last one value is only valid for a SLAP-like service, but not for a standard SLAP service. See Appendix B

- Exactly one field **SHOULD** have utype="**ssldm:Line.species.name**" with datatype="char", arraysize="*" and ucd="phys.atmol.element", containing a name of the chemical element source of this line.

- Exactly one field **SHOULD** have utype="**ssldm:Line.initialLevel.name**", with datatype="char", arraysize="*" and ucd="phys.atmol.initial;phys.atmol.level", containing a full description of the initial level of the transition originating the line.

- Exactly one field **SHOULD** have utype=”**ssldm:Line.finalLevel.name**”, with datatype="char", arraysize="*" and ucd=”phys.atmol.final;phys.atmol.level”, containing a full description of the final level of the transition originating the line.

- Exactly one field **MAY** have
  utype=”**ssldm:Line.observedWavelength.value**”,with datatype="double" and ucd="em.wl", containing the observed wavelength in the vacuum of the transition originating the line in meters, as it was observed. This may be used by observational spectral line databases.

- Exactly one field **MAY** have utype=”**slap:Query.Score**”, with datatype="double". A line with a higher score more closely matches the query parameters. The query response table should normally be returned sorted in order of decreasing values of score, with the top-scoring items at the top of the list. The details of the heuristic used to compute the score are left to the service. This is particularly useful for the theoretical spectral line databases examples explained in section 3. Please note that a reference to the relevant characterization namespace is needed in the VOTable response.

xmlns:slap ="http://www.ivoa.net/xml/SimpleLineAccessDM/SimpleLineAccessDM -v1.0.xsd"

- Exactly one field **MAY** have
  utype=”**ssldm:Line.initialLevel.energy.value**”, with datatype="double" and ucd="phys.energy;phys.atmol.initial;phys.atmol.level" and exactly one field **MAY** have utype=”**ssldm:Line.finalLevel.energy.value**”, with datatype="double" and ucd="phys.energy;phys.atmol.final;phys.atmol.level",
  describing the energy for the starting and final levels of the transition respectively which originates this line. To unify results, the value must appear in Joules.

  It is allowed to have more than one **Line.initialLevel.energy.value** and **Line.finalLevel.energy.value** fields in different units in order to preserve the precision of the original value prior to unit conversion in the client. If this is the case and to get backwards compatibility with already existing services, there MUST be one field with
  utype=”**ssldm:Line.initialLevel.energy.value**” and one with utype= **ssldm:Line.finalLevel.energy.value**” unit=”Joules” in the
  VOTable response. Other fields with the same utype should have a different value in the unit field descriptor.

- Exactly one field **MAY** have
  utype=”**ssldm:Line.initialLevel.configuration**” and one with

utype="**ssldm:Line.finalLevel.configuration**", with datatype="char" and arraysize="*" describing the electron configuration of the initial/final levels of the line.

The format of the string representing the electron configuration is as follows. For atomic levels:
Fe basic atomic level configuration is:

**Examples:**
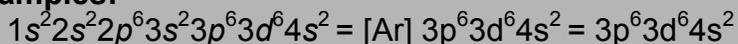

Where we have subtracted the closed shell configuration from the enumeration.

There is no required syntax for this string; however, it is recommended to use a LaTeX-styled [9] convention to describe sub-indexes, super-indexes, greek characters, etc. That means the previous atomic configuration could be serialized as string in the SLAP response in the following way:

**Examples:**
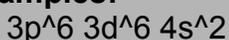

In the case of molecular levels, we usually need to use Greek symbols, so, e.g.,

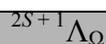

Would be serialized as

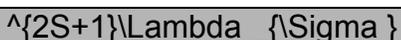

See [9] for a complete reference.

o Exactly one field **MAY** have utype="**ssldm:Line.initialLevel.quantumState**" and one with utype="**ssldm:Line.finalLevel. quantumState**", with datatype="char" and arraysize="*" describing the quantumState of the initial and final levels in a parseable string representation. The format must comply with the following syntax:

[label:type:numerator:denominator;label:type:numerator:denominator; …][…]

where:

- label is a string legal value for the model component given by
  **ssldm:Line.initialLevel.quantumState.quantumNumber.label**
- type is a string legal value for the model component given by
  **ssldm:Line.initialLevel.quantumState.quantumNumber.type**
- numerator is a integer legal value for the model component given by
  **ssldm:Line.initialLevel.quantumState.quantumNumber.numeratorValue**
- denominator is a integer legal value for the model component given  by
  **ssldm:Line.initialLevel.quantumState.quantumNumber.denominatorValue**

(Please refer to the Simple Spectral Line Data Model for meaning and format)
Quantum Numbers are separated by the ";" character.

To allow that a level needs more than one quantum state to be described, the
quantum states would be enclosed by square brackets.

---

**Example:**

[J:totalAngularMomentumJ:1:1;F1:totalAngularMomemtumF:0:1;
F:totalAngularMomemtum:1:1]

---

- o One field **MAY** have utype="**ssldm:Process.type**" with datatype="char"
  and arraysize="*"  identifying the types of physical processes responsible
  for the generation of the modification of its physical properties. As more
  than one value is possible per line, the values would be separated by the
  ":" character, in the following way:

---

**Example:**

   Matter-radiation interaction:Broadening

---

   Valid values are: "Matter-radiation interaction", 'Matter-matter interaction",
   "Energy shift", "Broadening".

- o One field **MAY** have utype="**ssldm:Process.name**" with datatype="char"
  and arraysize="*", identifying a description of the physical processes
  responsible for the generation of for the modification of its physical
  properties. As more than one value is possible per line, the values would
  be separated by the ":" character, in the following way:

---

**Example:**

   Photoionization:Natural broadening

---

This list should be in correspondence with the value of the possible process types described in previous section. Please notice that ":" is a reserved character.

o One or more fields **MAY** have ucd="meta.bib", with datatype="char" and arraysize="*", to specify an http link that contains a scientific publication related to the spectral line. Since the URL will often contain metacharacters, the URL is normally enclosed in an XML CDATA section (<![CDATA[...]]>) or otherwise encoded to escape any embedded metacharacters.

o One or more fields **MAY** have ucd="meta.ref.url" with datatype="char" and arraysize="*" and free names, to specify URLs that contains extra information related to the spectral line. Same criteria than before about the CDATA section.

In the case of observational line databases, some characterization information of the observation itself could be relevant. Next, we present some examples of possible observation-related output metadata (note that all the following fields are in line to the ones described in the SSAP specification)

o Exactly one field **MAY** have utype="**Target.Name**", with datatype="char", and arraysize="*" containing a short string identifying the observed astronomical object, suitable for input to a name resolver.

o Exactly one field **MAY** have utype="**char:SpatialAxis.Coverage.Location.Value**", with datatype="double", arraysize="*" and ucd="pos", containing the observation position of the observation in the format: ra dec, white space separated and both in deg. Please note that a reference to the relevant characterization namespace is needed in the VOTable response.

xmlns:char ="http://www.ivoa.net/xml/Characterisation/Characterisation-v1.11.xsd"

o Exactly one field **MAY** have utype="**char:TimeAxis.Coverage.Bounds.Start**", with datatype="char", arraysize="*" and ucd=" time.start;obs.exposure", containing the start time for the observation in MJD with units of days. Please note that a reference to the relevant characterization namespace is needed in the VOTable response.

o Exactly one field **MAY** have utype="**char:TimeAxis.Coverage.Bounds.Stop**", with datatype="char", arraysize="*" and ucd=" time.stop;obs.exposure", containing the end time for the observation in MJD with units of days. Please note that a reference to the relevant characterization namespace is needed in the VOTable response.

## 6.2 Non-Standard output fields

In many occasions, extra scientifically interesting parameters may be added to the output. Implementers are encouraged to add descriptions and UCDs to the return fields to clarify the meaning of this information and utypes to the Line Data Model or other existing IVOA Data Model, whenever possible.

# 7   Registration and discovering of SLAP resources

Users can discover a service instance through an IVOA-compliant registry [11]. The description of the instance in a registry is an XML document that complies with the VOResource resource description standard [12].  In particular, the description is recognized as an instance of a SLAP service when it complies with the SLAP VOResource extension [13].

# Appendix A: Range definition for input parameters

The Simple Image Access Protocol (SIAP), Simple Spectrum Access Protocol (SSAP) and the Simple Line Access Protocol (SLAP) are URL based protocols, i.e., the constraints in the queries for the input parameters need to be done in a valid URL.

However, the URL standard is not very specific defining special constraints as, e.g., ranges.  We describe here some examples of accepted range syntax for URL DAL protocols:

| param = x | Equality | Equivalent to param=x |
|---|---|---|
| param = x/y | Closed range | Equivalent to param>=X AND param<=y |
| param = x/ | Open range | Equivalent to param>=x |
| param = x,y | List | Equivalent to param=x OR param=y |
| param = x/a,b/y | Range list | Equivalent to [param<=x AND param<=a] OR [param>=b and param<=y] |

# Appendix B: SLAP services for uncorrected and unidentified lines

One important constraint for observational line databases in the SLAP definition is to limit the lines in the SLAP output to corrected lines, i.e. only the lines that have been corrected from wavelength shifts (including not only the astrophysical redshift, but also any process that could produce a line peak displacement). (Note that unidentified lines can be distributed as long as these lines could be corrected in wavelength. See observational "**ssldm:Line.title**" recommended syntax.)

The reason for that is, if a user wants to compare two different SLAP service outputs or to compare one SLAP service output to one SSAP service output, the data comparison requires previous correction. This correction is often quite complex and it should be done by the data providers, the ones who have the knowledge and expertise to correct the data properly, and not by a client application, as this process is prone to scientific errors.

Even if this kind of SLAP-like service cannot be registered in the IVOA as standard SLAPs, the creation of services for internal project consumption could be desirable.

If a service is able to provide both corrected and uncorrected lines, the service could be registered as soon as a default call to the service only returns the corrected lines as output. The following optional parameter could be used to select other types of lines.

o **CORRECTION_STATUS**
A service **MAY** have a search parameter called CORRECTION_STATUS. This parameter would constraint the search to lines with a certain level of identification. As specified in [1] the possible values are:

UNCORRECTED
CORRECTED

And the default value MUST be "CORRECTED".
(ucd="em.line;meta.id.cross";utype="**ssldm:Line.correctionStatus"**)

o **IDENTIFICATION_STATUS**
A service **MAY** have a search parameter called IDENTIFICATION_STATUS. This parameter would constrain the search to lines with a certain level of identification. As specified in [1] the possible values are:

UNIDENTIFIED
IDENTIFICATION_UNCERTAIN
IDENTIFICATION_PROVISIONAL
IDENTIFIED

(ucd="em.line;meta.id.cross";utype="**ssldm:Line.identificationStatus"**)

For theoretical line databases, identified will be synonymous to predicted.

For the output:

- o Exactly one field **MAY** have utype="**ssldm:Line.correctionStatus**", with datatype="char" and ucd="em.wl.central;meta.id.cross", describing if the "ldm:Line.wavelength" has been calculated.

  Please note that for uncorrected lines the compulsory field with utype="**ldm:Line.wavelength**" (corrected wavelength in vacuum) will contain either the observed wavelength for "Uncorrected" or a provisional assignment value.

- o Exactly one field **MAY** have utype="**ssldm:Line.identificationStatus**", with datatype="char" and ucd="em.line;meta.id.cross", describing the identification status of the line.

# Appendix C: SLAP valid response example

```
<?xml version="1.0" encoding="UTF-8"?>
<VOTABLE xmlns:xsi="http://www.w3.org/2001/XMLSchema-instance"
xsi:noNamespaceSchemaLocation="xmlns:http://www.ivoa.net/xml/VOTable/VOTable-1.1.xsd"
xmlns:ssldm ="http://www.ivoa.net/xml/SimpleSpectrumLineDM/SimpleSpectrumLineDM-
v1.0.xsd" version="1.0">
<RESOURCE type="results">
<DESCRIPTION>European Space Astronomy Centre - Simple Line Access Protocol
(SLAP)</DESCRIPTION>
<INFO name="QUERY_STATUS" value="OK"/>
<INFO name="SERVICE_PROTOCOL" value="1.0">SLAP</INFO>
<INFO name="REQUEST" value="queryData"/>
<INFO name=" WAVELENGTH" value="7.6E-6/1.E-5"/>

<TABLE>
<FIELD ucd="meta.id;obs" name="OBSNO" datatype="char" arraysize="*"/>
<FIELD ucd="em.wl" name="WAVELENGTH" utype="ssldm:Line.wavelength.value"
datatype="double"/>
<FIELD ucd="em.freq" name="FREQUENCY" utype=" ssldm:Line.frequency.value"
datatype="double"/>
<FIELD ucd="em.wavenumber" name="WAVE_NUMBER"
utype="ssldm:Line.wavenumber.value" datatype="double"/>
<FIELD ucd="spect.line.width" name="LINEWIDTH"
utype="ssldm:Line.observedBroadeningCoefficient.value" datatype="double"/>
<FIELD ucd="spect.line.width;meta.unit" name="LINEWIDTH_UNIT" datatype="char"
arraysize="*"/>
<FIELD ucd="meta.bib" name="ADS_CODE" datatype="char" arraysize="*"/>
```

```xml
<FIELD ucd="em.line" name="IDENTIFICATION" utype="ldm:Line.title" datatype="char"
arraysize="*"/>
<FIELD ucd="phys.atmol.transition" name="TRANSITION" datatype="char" arraysize="*"/>
<FIELD name="LINE_TYPE" datatype="char" arraysize="*"/>
<FIELD ucd="spect.line;phot.flux" name="FLUX" utype="ssldm:Line.observedFlux.value"
datatype="double"/>
<FIELD ucd="spect.line.intensity" name="PEAK_INTENSITY"
utype="ssldm:Line.observedIntensity.value" datatype="double"/>
<FIELD ucd="spect.line.intensity;stat.snr" name="SIGNAL_TO_NOISE" utype="
ssldm:Line.significanceOfDetection.value" datatype="double"/>
<DATA>
<TABLEDATA>

<TR>
<TD>116003190</TD>
<TD>8.03E-6</TD>
<TD>37333.748443</TD>
<TD>1245.330012</TD>
<TD>0.008580</TD>
<TD>micron</TD>
<TD>2001ApJ...552..544F</TD>
<TD>H2</TD>
<TD>0-0 S(4) </TD>
<TD>L</TD>
<TD>6.800000088194428E-16</TD>
<TD> </TD>
<TD> </TD>
</TR>

<TR>
<TD>116003190</TD>
<TD>8.997E-6</TD>
<TD>33321.107035</TD>
<TD>1111.481604</TD>
<TD>0.010960</TD>
<TD>micron        </TD>
<TD>2001ApJ...552..544F</TD>
<TD>[ArIII] </TD>
<TD>3P2-3P1</TD>
<TD>L</TD>
<TD>4.8899999785631705E-15</TD>
<TD> </TD>
<TD>0.000000</TD>
</TR>

….. more lines data……

</TABLEDATA>
</DATA>
</TABLE>
</RESOURCE>
</VOTABLE>
```

# Appendix D: SLAP data model summary

| UTYPE | UCD | Description | Data Type | Array Size |
|---|---|---|---|---|
| **ssldm:Line.title (REQUIRED)** | em.line | small description identifying the line | char | * |
| **ssldm:Line.wavelength.value**<br><br>**(REQUIRED)** | em.wl | wavelength in the vacuum of the transition originating the line in meters. | char | * |
| **ssldm:Line.initialLevel.energy.value** | phys.energy; phys.atmol.i nitial;phys.at mol.level | energy for the INITIAL level of the transition | double | |
| **ssldm:Line.finalLevel.energy.value** | phys.energy; phys.atmol.fi nal;phys.atm ol.level | energy for the FINAL level of the transition | double | |
| **ssldm:Line.environment.temperature.value** | phys.temperat ure | expected temperature of the object | double | |
| **ssldm:Line.einsteinA.value** | phys.atmol.tra nsProb | Einstein A coefficient, probability per unit time s$^{-1}$ for spontaneous emission in a bound-bound transition | double | |
| **ssldm:Process.type** | meta.title | physical process type responsible for the generation of the line or for the modification of its physical properties | char | * |
| **ssldm:Process.name** | meta.title | physical process exact description responsible for the generation of the line or for the modification of its physical properties | char | * |
| **ssldm:Line.identificationStatus** | | identification status of the line | char | * |
| **ssldm:Line.initialLevel.name** | phys.atmol.i nitial;phys.at mol.level | description of the initial level of the transition originating the line | char | * |
| **ssldm:Line.finalLevel.name** | phys.atmol.fi nal;phys.atm ol.level | description of the final level of the transition originating the line | char | * |
| **ssldm:Line.observedWavelength.value** | em.wl | observed wavelength in the vacuum of the transition originating the line in meters | char | * |
| **slap:Query.Score** | | Line ranking "more closely matches the query parameters" | double | |
| **ssldm:Line.initialLevel.configuration** | phys.atmol.co nfiguration | Description of the electron configuration of the initial level of | char | * |

| | | the line | | |
|---|---|---|---|---|
| ssldm:Line.finalLevel.configuration | phys.atmol.configuration | describing the electron configuration of the final level of the line | char | * |
| ssldm:Line.initialLevel.quantumState | meta.title | Description of the quantum state of the initial level in a parseable string representation | char | * |
| ssldm:Line.finalLevel. quantumState | meta.title | Description of the quantum state of the final level in a parseable string representation | char | * |
| Target.Name | meta.id;src | short string identifying the observed astronomical object, suitable for input to a name resolver | char | * |
| char:SpatialAxis.Coverage.Location.Value | Pos | observation position of the observation in the format: ra dec, white space separated and both in deg | char | * |
| char:TimeAxis.Coverage.Bounds.Start | time.start;obs.exposure | start time for the observation in MJD with units of days | char | * |
| char:TimeAxis.Coverage.Bounds.Stop | time.stop;obs.exposure | end time for the observation in MJD with units of days | char | * |

Please note that only the two first utypes correspond to required fields in the protocol.

# References

 **[1] [Osuna/Guainazzi/Salgado/Dubernet/Roueff]**
Simple Spectral Lines Data Model
http://www.ivoa.net/Documents/SSLDM/

**[2] [Tody/Plante]**
Simple Image Access Specification,
http://www.ivoa.net/Documents/latest/SIA.html

**[3] [Hanisch/Arviset/Genova/Rino]**
IVOA Document Standards,
http://www.ivoa.net/Documents/DocStd/20090302/PR-DocStd-1.2-20090302.html

**[4] [Smith/Heise/Esmond/Kurucz]**
Atomic spectral line database from CD-ROM 23 of R. L. Kurucz.,
http://cfa-www.harvard.edu/amdata/ampdata/kurucz23/sekur.html

**[5] [Various]**
NIST National Institute of Standards and Technology,
NIST Atomic Spectra Database Lines,
http://physics.nist.gov/PhysRefData/ASD/lines_form.html